# Minimum clique partition in unit disk graphs

Adrian Dumitrescu*   János Pach†

November 4, 2018

## Abstract

The minimum clique partition (MCP) problem is that of partitioning the vertex set of a given graph into a minimum number of cliques. Given $n$ points in the plane, the corresponding unit disk graph (UDG) has these points as vertices, and edges connecting points at distance at most 1. MCP in unit disk graphs is known to be NP-hard and several constant factor approximations are known, including a recent PTAS. We present two improved approximation algorithms for minimum clique partition in unit disk graphs:

(I) A polynomial time approximation scheme (PTAS) running in time $n^{O(1/\varepsilon^2)}$. This improves on a previous PTAS with $n^{O(1/\varepsilon^4)}$ running time [22].

(II) A randomized quadratic-time algorithm with approximation ratio 2.16. This improves on a ratio 3 algorithm with $O(n^2)$ running time [7].

## 1 Introduction

*Unit disk graphs* (UDGs) can be defined in three equivalent ways, as follows [8]: (i) For $n$ points in the plane, form a graph with $n$ vertices corresponding to the $n$ points, and an edge between two vertices if and only if the distance between the two points is at most 1. (ii) For $n$ unit circles in the plane, form a graph with $n$ vertices corresponding to the $n$ circles, and an edge between two vertices if and only if one of the corresponding circles contains the other's center. (iii) For $n$ circles of unit diameter in the plane, form a graph with $n$ vertices corresponding to the $n$ circles, and an edge between two vertices if and only if the two circles intersect. If in definition (iii) above, the (interiors of the) circles are non-overlapping, the unit disk graph is called a *unit coin graph*, see [7]. The three models for the above definitions are known as: (i) the *proximity model*, (ii) the *containment model*, and (iii) the *intersection model*.

Most applications of unit disk graphs arise in wireless network research: usually, two nodes can directly communicate if they lie in the unit disks placed at each others' centers. There is a vast literature on algorithmic problems studied on UDGs, here we mention only a few: [1, 7, 8, 10, 11, 18, 20, 22]. A recent survey is [2]. The seminal paper by Clark et al. [8] studies of complexities of several classical graph optimization problems, known to be NP-complete in general, in the setting of unit disk graphs: coloring, clique, independent set, vertex cover, domination, independent domination, and connected domination. For instance, it is shown there [8] that in unit disk graphs a maximum clique can be found in polynomial time. On the other hand, some closely related problems, such as finding a proper 3-coloring, remain NP-complete even in this setting [8]. Breu and Kirkpatrick [4] later showed that the problem of recognizing unit disk graphs is NP-hard, and thereby answered one of problems left open in [8].

*Department of Computer Science, University of Wisconsin–Milwaukee, USA. Email: `ad@cs.uwm.edu`.
†Ecole Polytechnique Fédérale de Lausanne and City College, New York. Email: `pach@cims.nyu.edu`.

Given a set $S$ of $n$ points in the plane, a partition of $S$ into $q$ sets $C_1, C_2, \ldots, C_q$ is called a *q-clustering*, and the individual sets are called its *clusters*; see [5]. Feder and Greene [12] have shown that it is NP-hard to find a $q$-clustering in which the maximum diameter of the clusters is within a factor of 1.97 of the optimum. Clustering is a broad area in graph optimization, that is particularly relevant to wireless networks: typically one wants to find large groups that can mutually communicate with each other. Mutual *proximity* of nodes in a cluster is therefore a key criterion, and *cliques* as clusters describe it the best. The *minimum clique partition* (MCP) problem is that of partitioning the vertex set of a given graph into a minimum number of cliques. For general graphs, MCP is equivalent to minimum graph coloring of the complement graph, which is known to be inapproximable within $n^{1-\varepsilon}$, for any $\varepsilon > 0$, unless P=NP [24]. The best known approximation algorithm for the graph coloring problem is an $O(n \frac{(\log \log n)^2}{(\log n)^3})$-approximation, due to Halldórsson [15].

Supowit has shown that the MCP problem in UDGs is NP-complete [23]. Cerioli et al. [7] have shown that the problem remains NP-complete even for unit coin graphs. They also gave a 3-approximation algorithm for MCP in UDGs running in quadratic time. Recently, Pirwani and Salavatipour [22] have devised a PTAS for MCP in unit disk graphs, that relies on the separability property of an optimal clique partition, first established by Capoyleas et al. [5] in the early 1990s. The authors of [22] rediscovered this property, apparently being unaware of the old solution.

**Theorem 1.** (Capoyleas, Rote and Woeginger [5]). *Given a finite point set $S$ in the plane, there exists an optimal clique partition where the convex hulls of the cliques are non-overlapping.*

**Our Results.** By refining the ideas used in [5] and [22], in Section 2 we obtain an $n^{O(1/\varepsilon^2)}$ (polynomial) time approximation scheme (PTAS) for MCP. This improves on the previously best algorithm, which has running time $n^{O(1/\varepsilon^4)}$ [22].

**Theorem 2.** *There exists a PTAS for MCP in unit disk graphs, which computes an $(1 + \varepsilon)$-approximation in $n^{O(1/\varepsilon^2)}$ time.*

In Section 3, we revisit the ratio 3 approximation algorithm of Cerioli et al. [7], which builds on the ideas of Breu [3]. The ratio 3 approximation was the best achievable with a practical algorithm, running in $O(n^2)$ time. Based on their idea, we show that a randomized variant of the algorithm computes a 2.16-approximate solution within about the same running time.

**Theorem 3.** *For any $0 < \varepsilon, \delta < 1$, there is an algorithm for MCP in unit disk graphs, which computes a solution at most $1 + 2/\sqrt{3} + \varepsilon$ times the optimal with probability at least $1 - \delta$ in $O\left(\frac{1}{\varepsilon} \cdot \ln \frac{1}{\delta} \cdot n^2\right)$ time. In particular, a $\frac{181}{84}$-approximate solution can be computed with probability at least $1 - \delta$ in $O(\ln \frac{1}{\delta} \cdot n^2)$ time.*

In Section 4, we give a streamlined proof of Theorem 1, which makes our paper self-contained.

**Definitions and notations.** We say that a set $S$ of points in the plane is in *general position* if no three of its points are collinear. For a point set $S$, let conv$(S)$ be the convex hull of $S$. If there is no danger of confusion, we use the same symbol to sometimes denote a polygonal region (or, simply, polygon) and its boundary. Given a convex polygon $P$, let $V(P)$ denote its vertex set. For a given set $S$ of input points, let $G = G(S)$ denote the UDG defined by the set of unit disks centered at the elements of $S$, and let $z = z(S)$ denote the minimum number of cliques in a clique partition of $G$, that is, the solution of the MCP problem for $S$ (and for $G = G(S)$).

Two convex polygons $A$ and $B$ in the plane are said to be *overlapping* if $\text{area}(A \cap B) > 0$, and *non-overlapping* otherwise. Analogously, we say that two cliques in some clique partition are overlapping or non-overlapping.

## 2 A faster $n^{O(1/\varepsilon^2)}$-time PTAS

In this section, we prove Theorem 2. First notice that Theorem 1 holds in a slightly stronger form: there exists an optimal clique partition in which the (convex hulls) of the cliques are *pairwise disjoint*, rather than simply non-overlapping. For any two cliques whose convex hulls share points from $S$ on two of their boundary edges, these points can be arbitrarily assigned to one of the cliques. By repeatedly applying this operation, one can obtain a clique partition, where the convex hulls of the cliques are pairwise disjoint. This operation is not necessary if the points are assumed to be in general position. We start with two simple facts (the first one is well-known):

**Lemma 1.** *Given two disjoint convex polygons $P$ and $Q$ in the plane, there exists a separating (tangent) line determined by a pair of vertices in $V(P) \cup V(Q)$.*

**Lemma 2.** *The size of an optimal solution for MCP for a set of points in a $k \times k$ square cell $\tau$ is at most $2k^2 + 3k$.*

*Proof.* Cover a $k \times k$ grid cell $\tau$ by smaller square cells of size $\frac{1}{\sqrt{2}} \times \frac{1}{\sqrt{2}}$. Recall that $k \in \mathbb{N}$, and $k \geq 6$. The number of smaller cells is at most $\lceil k\sqrt{2} \rceil^2 \leq (k\sqrt{2} + 1)^2 \leq 2k^2 + 3k$. Since any set of points inside a $\frac{1}{\sqrt{2}} \times \frac{1}{\sqrt{2}}$ square forms a clique, it follows that an optimal clique partition consists of at most $2k^2 + 3k$ cliques. □

Our PTAS, which follows the same overall structure of the PTAS in [22], is outlined below. Our improvement in the running time comes from speeding up STEP 2.

STEP 1. Partition the region of the plane containing $S$ using a randomly shifted grid whose cell size is $k \times k$, with $k = \lceil \frac{16}{\varepsilon} \rceil$. If $k$ is large enough, the probability that any fixed clique in an optimum clique partition is cut by this grid (and therefore belongs to two or more grid cells) is small, namely $O(1/k)$.

STEP 2. Compute an optimal clique partition in each cell $\tau$ of the grid and return the union of these cliques. The expected size of the solution is at most $(1 + \epsilon)z$.

By repeating these two steps in $O(\ln \frac{1}{\delta})$ independent random trials, the solution is at most $(1 + \epsilon)z$ with probability at least $1 - \delta$, for any prescribed $0 < \delta < 1$.

We now show that STEP 2 can be performed in $n^{O(1/\varepsilon^2)}$ time. For simplicity, we still use $n$ to denote the number of points of $S$ in a fixed grid cell $\tau$. It suffices to show that an optimal solution in $\tau$ can be computed in $n^{O(1/\varepsilon^2)}$ time, since the overall running time is then bounded by the same expression. The details are as follows.

Let $\tau$ be a fixed grid cell. Let $q_{\max} = 2k^2 + 3k$ be the upper bound from Lemma 2. For $q = 1, 2, \ldots, q_{\max}$, the algorithm checks whether there exists a clique partition of size $q$ for the points in $S \cap \tau$. Assume that $q$ is the optimal size. Consider an (unknown) optimal clique partition with pairwise disjoint cliques $C_1, \ldots, C_q \subset S \cap \tau$. Write $\mathcal{C}_\tau = \{C_1, \ldots, C_q\}$. Assume that we are given $q$ *representative points* $r_1, \ldots, r_q \in S \cap \tau$, where $r_i \in C_i$. Define the *proximity graph* $X = (R, F)$ with vertex set $R = \{r_1, \ldots, r_q\}$, where $r_i r_j \in F$ if and only if $|r_i r_j| \leq 2$. Two key steps are the following:

**Lemma 3.** *If $r_i r_j \notin F$, and $p \in S \cap \tau$ satisfies $|pr_i| \leq 1$, then $p \notin C_j$.*

*Proof.* Assuming $p \in C_j$ yields $|pr_j| \leq 1$. By the triangle inequality, we have $|r_i r_j| \leq |pr_i| + |pr_j| \leq 2$, which contradicts the assumption $r_i r_j \notin F$ in the lemma. □

**Lemma 4.** *The maximum degree in $X$ is at most 79.*

*Proof.* Let $r_i$ be an arbitrary vertex of $X$, and let $N(r_i)$ be the set of vertices connected to $r_i$ by an edge in $X$. Note that all cliques in $\mathcal{C}_\tau$ whose representative points are in $\{r_i\} \cup N(r_i)$ are contained in the circle $\Omega$ of radius 3 centered at $r_i$. For a short argument, observe that the axis-aligned square circumscribed around $\Omega$ has side length 6, and recall that $\mathcal{C}_\tau$ is a minimum clique for the points in $S \cap \tau$. Cover $\tau$ by smaller rectangular cells of size $\frac{3}{5} \times \frac{4}{5}$, aligned with $\tau$. The number of smaller cells is at most $10 \cdot 8 = 80$. Since any set of points inside a $\frac{3}{5} \times \frac{4}{5}$ rectangle forms a clique, it follows that $\mathcal{C}_\tau$ consists of at most 80 cliques. Hence $|N(r_i)| \leq 79$, so the maximum degree in $X$ is at most 79, as claimed. □

Note that, by Lemma 1, for any pair of representatives $r_i, r_j \in R$, there exists a line separating $C_i$ from $C_j$, that passes through two points of $V(\text{conv}(C_i)) \cup V(\text{conv}(C_j))$. In particular, this holds for pairs $r_i, r_j \in R$, such that $r_i r_j \in F$ (that is, pairs in the proximity graph). Refer to Fig. 1.

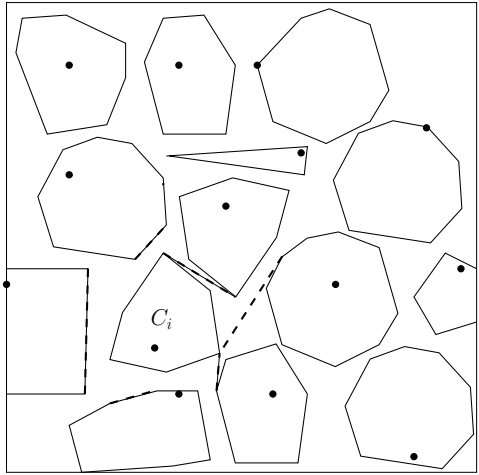

Figure 1: A clique partition in cell $\tau$, with pairwise disjoint cliques, drawn as convex polygons. Lines separating $C_i$ from other cliques $C_j$, where $r_i r_j \in F$ are drawn with dashed lines. Representative points in each clique are drawn as filled circles.

Assume that we are given: (i) $q$ representative points $r_1, \ldots, r_q \in S \cap \tau$, where $r_i \in C_i$; (ii) the proximity graph $X$ with vertex set $R = \{r_1, \ldots, r_q\}$, and (iii) the pairs of vertices (points in $S \cap \tau$) incident to each of the separating lines for the pairs $r_i r_j \in F$.

Having all the above information, its validity can be verified in $O(n^2)$ time, that is, if it corresponds to a valid clique partition of the points in $S \cap \tau$: more precisely, for each representative point $r_i$, the points in $S$ at distance at most 1 from $r_i$, and that lie on the same side as $r_i$ of the separating lines for the pairs $C_i, C_j$ corresponding to the edges $r_i r_j \in F$, must form a clique.

Our algorithm simply generates all guesses of the above form, as specified by (i), (ii), and (iii), and checks their validity. If at least one valid clique partition of size $q$ for the points in $S \cap \tau$, is found, the algorithm terminates, since there is no need to try the larger values of $q$. For discussing the time complexity of the algorithm, we take into account the following elements:

1. There are at most $\binom{n}{q} \leq n^q$ choices for the representative points.

2. Since the maximum degree in $X$ is at most 79, the number of edges of $X$ is at most $79q/2 = 39.5q$, and the total number of proximity graphs on vertices $r_1, \ldots, r_q$, is at most $2^{39.5q \log q + O(q)}$.

3. Given $X$, there are at most $\binom{n}{79q} \leq n^{79q}/(79q)!$ choices for the separating lines: two endpoint choices per separating segment.

4. Given $X$, say with $m \leq 79q/2$ edges, and a set of $m$ separating lines, there are at most $(39.5q)! = 2^{39.5q \log q + O(q)}$ perfect matchings between them.

Recall that $q \leq q_{\max} = 2k^2 + 3k$, where $k = O(\frac{1}{\varepsilon})$. We obtain the following overall upper bound on the running time:
$$2^{O(q)} \cdot n^{80q} = 2^{O(k^2)} \cdot n^{160k^2 + O(k)} = n^{O(1/\varepsilon^2)}.$$

The proof of Theorem 2 is now complete. □

**Remark 1.** For simplicity of the arguments and of the resulting bounds, we have often used only rough estimates. For instance, in Lemma 2, one can get a better bound by using hexagonal cells, and the degree bound in Lemma 4 can also be reduced by a more careful argument. The same applies to some of the bounds used in [22]; for example, in the analysis of STEP 1 of the PTAS from [22], for a given clique partition of $G$, the probability that a clique is cut by both a vertical and a horizontal line of the grid is only about $O(\frac{1}{k^2})$ rather than $O(\frac{1}{k})$. So most of the cliques are cut in two rather than four parts by the randomly shifted grid. The bottom line is that for obtaining a practical PTAS, many theoretical aspects and implementation details need to be addressed altogether at the same time.

**Remark 2.** Capoyleas et al. [5] proposed a different approach for computing a MCP in a UDG, under the assumption that the number of cliques in the optimal partition is bounded by a constant. Their proof is based on the following lemma of L. Fejes Tóth [13] (see also Cassels [6], Heppes [17], Edelsbrunner et al. [9], or [21, p. 23]): Given a collection of pairwise disjoint closed convex sets $C_1, \ldots, C_q$ in a square $Q$, say, there exists a subdivision of $Q$ into convex pieces with a total number of at most $6q$ sides such that each piece contains at most one $C_i$.

Let $C_i'$ denote the piece containing $C_i$ in such a subdivision ($1 \leq i \leq q$). Define a graph $G^*$ on the vertex set $\{C_1', \ldots, C_q'\}$ by connecting $C_i'$ and $C_j'$ with an edge if and only if they share a boundary segment. In the proof of their Theorem 9, Capoyleas et al. applied Fejes Tóth's lemma to the convex hulls $C_i$ of $q$ cliques that form a clique partition, and suggested that a MCP can be obtained by enumerating all possible planar graphs $G^*$ on $q$ vertices and trying all possible connecting lines of the $n$ input points, as potential sides of the convex polygonal pieces $C_i'$. Since it is not at all obvious (to us) whether there exists a subdivision in which all sides belong to such connecting lines, in our proof of Theorem 2, we decided to follow a different approach.

## 3 A randomized $2.16$-approximation algorithm

In this section we prove Theorem 3. Following [3], a unit disk graph $G$ is called a $\tau$-*strip* graph, if all the points lie in a parallel strip of width $\tau$. As shown by Breu in his PhD thesis, a $\tau$-strip graph for $\tau \leq \sqrt{3}/2$ is the complement of a comparability graph, therefore the minimum clique partition problem can be solved (exactly) in polynomial time on such strip graphs [3]. Specifically this amounts to coloring the complement $\overline{G}$ of $G$, and this takes $O(n^2)$ time [14].

A 3-approximation algorithm for MCP based on this idea is given in [7]: it partitions the plane (the region containing all the points) into parallel (say, horizontal) strips of width $\sqrt{3}/2$, then computes an optimal clique partition in each strip, and finally outputs the union of all the cliques computed. The algorithm runs in $O(n^2)$ time. Instead of strips of an irrational width, $\sqrt{3}/2$, one can use slightly thinner strips, say of width $0.8$, and the same result still holds.

Here we present and analyze a randomized variant which uses strips of width $a = \sqrt{3}/2$. The system of strips is determined by placing a horizontal line at a random $y$-coordinate uniformly chosen in the interval $[0, a)$. In fact, the strips can be slightly thinner, so that their width is a rational number, see below.

**Proof of Theorem 3.** Let $\xi = 1 + \frac{1}{a}$, where $a = \sqrt{3}/2$. Let $OPT = C_1 \cup \ldots C_z$ be an optimal clique partition. Obviously, we have $z \leq n$. For $1 \leq i \leq z$, let $X_i$ be the number of parts in which $C_i$ is split by the system of horizontal strips, and let $b_i$ be the vertical width of $C_i$. Obviously $b_i \leq 1$, and $X_i \in \{1, 2, 3\}$. Let $X$ be the number of cliques output by the algorithm. Clearly, $X = \sum_{i=1}^{z} X_i$ and $z \leq X \leq 3z$. We distinguish two cases:

*Case 1:* $b_i \leq a$. Then $X_i \in \{1, 2\}$, thus $\mathrm{E}[X_i] \leq 2 < 1 + 1/a$.

*Case 2:* $b_i > a$. Then $X_i \in \{1, 2, 3\}$. It is easy to see that

$$\mathrm{Prob}[X_i = 3] = \frac{b_i}{a} - 1, \quad \text{and} \quad \mathrm{Prob}[X_i \leq 2] = 2 - \frac{b_i}{a},$$

thus

$$\mathrm{E}[X_i] \leq 2\left(2 - \frac{b_i}{a}\right) + 3\left(\frac{b_i}{a} - 1\right) = 1 + \frac{b_i}{a} \leq 1 + \frac{1}{a}.$$

Thus in both cases we have $\mathrm{E}[X_i] \leq 1 + \frac{1}{a}$. It follows immediately by the linearity of expectation that

$$\mathrm{E}[X] = \sum_{i=1}^{z} \mathrm{E}[X_i] \leq \left(1 + \frac{1}{a}\right) z = \xi z. \tag{1}$$

Observe that

$$\left(1 + \frac{\varepsilon}{3}\right) \xi \leq \xi + \varepsilon.$$

By Markov's inequality [19], the probability that the solution output by the algorithm after one round is larger than $(\xi + \varepsilon)z$, is bounded by:

$$\mathrm{Prob}[X \geq (\xi + \varepsilon)z] \leq \mathrm{Prob}\left[X \geq \left(1 + \frac{\varepsilon}{3}\right)\xi z\right] \leq \mathrm{Prob}\left[X \geq \left(1 + \frac{\varepsilon}{3}\right)\mathrm{E}[X]\right] \leq \frac{1}{1 + \frac{\varepsilon}{3}}. \tag{2}$$

Since the rounds are independent, it follows that

$$\mathrm{Prob}[X \geq (\xi + \varepsilon)z \text{ in } j \text{ rounds}] \leq \left(\frac{1}{1 + \frac{\varepsilon}{3}}\right)^j. \tag{3}$$

Recall the standard inequality $\ln(1 + x) \geq 0.9x$, for $0 \leq x \leq 0.1$. To make the above bound smaller than $\delta$, it is enough to choose

$$j \geq \frac{\ln \frac{1}{\delta}}{\ln\left(1 + \frac{\varepsilon}{3}\right)}, \quad \text{e.g.,} \quad j = \left\lceil \frac{\ln \frac{1}{\delta}}{0.3\varepsilon} \right\rceil. \tag{4}$$

The resulting running time of the algorithm is $O\left(\frac{1}{\varepsilon} \cdot \ln \frac{1}{\delta} \cdot n^2\right)$, as claimed.

Note that having strips of width $a = \sqrt{3}/2$ may be impractical, or may require additional complications, since $a \notin \mathbb{Q}$. Then one may use strips of slightly smaller width. We analyze this option below. First observe that if $0.8 \leq d \leq a = 0.866\ldots$, and a system of strips of width $d < a$ is chosen, the same algorithm can be used. Equation (1) becomes in this case:

$$\mathrm{E}[X] = \sum_{i=1}^{z} \mathrm{E}[X_i] \leq \left(1 + \frac{1}{d}\right) z. \tag{5}$$

We use standard facts from number theory pertaining to approximations of irrational numbers by continued fractions [16, Chapters X, XI]. Let $\xi = 1 + \frac{1}{a}$, and $r = p/q > \xi$ be a rational approximation of $\xi$ from above. We will discuss shortly how a suitable such approximation can be found. The algorithm randomly selects a strip partition, with strips of width $d$, where $d$ is the solution of

$$1 + \frac{1}{d} = \frac{p}{q}, \text{ namely } d = \frac{q}{p-q}.$$

Here $p$ and $q$, thus also $d$ will depend on the given $\varepsilon > 0$, and next we discuss their selection. It is easy to see that $\xi$ satisfies the quadratic equation

$$\xi = 2 + \frac{1}{6 + \frac{1}{\xi}},$$

thus its continued fraction expansion is periodic [16, pp. 143]:

$$\xi = [2, 6, 2, 6, 2, 6, \ldots] = [\dot{2}, \dot{6}].$$

Write

$$\xi_t = \frac{p_t}{q_t}, \quad t = 0, 1, 2, \ldots$$

for the convergents in the continued fraction expansion of $\xi$, obtained by Euclid's algorithm. For instance, the first four convergents of $\xi$ are:

$$\frac{p_0}{q_0} = \frac{2}{1}, \quad \frac{p_1}{q_1} = \frac{13}{6}, \quad \frac{p_2}{q_2} = \frac{28}{13}, \quad \frac{p_3}{q_3} = \frac{181}{84}.$$

By [16, Theorems 152,153,154,164], the convergents tend in the limit to $\xi$, and satisfy

$$\left|\frac{p_t}{q_t} - \xi\right| < \frac{1}{q_t^2},$$

with the odd convergents being strictly greater than $\xi$, decreasing to $\xi$ in the limit, and satisfying the recurrence

$$\frac{p_t}{q_t} = \frac{13 p_{t-2} + 2 q_{t-2}}{6 p_{t-2} + q_{t-2}}, \quad (t \text{ odd}).$$

Fix now $t > 0$ to be the smallest odd positive integer for which

$$q_t^2 \geq \left\lceil \frac{3}{\varepsilon} \right\rceil,$$

and set $p = p_t$, and $q = q_t$. By this choice,

$$\left|\frac{p_t}{q_t} - \xi\right| < \frac{1}{q_t^2} \leq \frac{\varepsilon}{3}, \text{ hence } \frac{p}{q} \leq \xi + \frac{\varepsilon}{3}.$$

By substituting this bound in Equation (5), we obtain that

$$\mathrm{E}[X] \leq \frac{p}{q} z \leq \left(\xi + \frac{\varepsilon}{3}\right) z. \tag{6}$$

Note also that

$$\left(\xi + \frac{\varepsilon}{3}\right)\left(1 + \frac{\varepsilon}{4}\right) \leq \xi + \varepsilon, \quad \text{for } \varepsilon \leq 1.$$

Combining the previous bounds, we can obtain the analogues of (2) and (3):

$$\begin{aligned}
\mathrm{Prob}[X \geq (\xi + \varepsilon)z] &\leq \mathrm{Prob}[X \geq \left(1 + \frac{\varepsilon}{4}\right)\left(\xi + \frac{\varepsilon}{3}\right)z] \\
&\leq \mathrm{Prob}[X \geq \left(1 + \frac{\varepsilon}{4}\right)\mathrm{E}[X]] \\
&\leq \frac{1}{1 + \frac{\varepsilon}{4}}.
\end{aligned} \tag{7}$$

$$\mathrm{Prob}[X \geq (\xi + \varepsilon)z \text{ in } j \text{ rounds}] \leq \left(\frac{1}{1 + \frac{\varepsilon}{4}}\right)^j. \tag{8}$$

As in Equation (4), we can make ensure that the error probability is at most $\delta$, by setting $j = \left\lceil \frac{\ln \frac{1}{\delta}}{0.225\varepsilon} \right\rceil$.

The special case with ratio $\frac{181}{84}$ mentioned in the theorem corresponds to the second odd convergent ($t = 3$). It uses strips of width $\frac{84}{97}$, and already gives $\varepsilon < 10^{-4}$. This completes the proof of Theorem 3. □

## 4   A short proof of Theorem 1

Let $\mathcal{C}$ be a clique partition for point set $S$. As in [5] (or in [22]), define an appropriate "potential function", $\Psi$, over the cliques in $\mathcal{C}$. The value of $\Psi$ for $\mathcal{C}$ equals the sum of the perimeters of the convex hulls of the cliques in $\mathcal{C}$:

$$\Psi(\mathcal{C}) = \sum_{C \in \mathcal{C}} \mathrm{per}(\mathrm{conv}(C)).$$

Let $\mathcal{C}$ be a clique partition minimizing $\Psi(\mathcal{C})$ and assume for contradiction that $\mathcal{C}$ has overlapping cliques. Let $C, D \in \mathcal{C}$ be a pair of overlapping cliques, and let $P = \mathrm{conv}(C)$, and $Q = \mathrm{conv}(D)$. Let $I = P \cap Q$, and $L = P \cup Q$. By definition, $I$ is a convex polygon with non-zero area. Let $P \setminus Q = \{X_0, \ldots, X_{k-1}\}$ be the set of connected regions of $P \setminus Q$ in clockwise order. Similarly, let $B \setminus A = \{Y_0, \ldots, Y_{k-1}\}$ be the set of connected regions of $Q \setminus P$ in clockwise order, where $X_0, Y_0$ are consecutive in this order. Each region $X_i$, or $Y_j$ is referred to as a *petal* (of $P$, or $Q$, respectively; see Fig. 2(left). Obviously, the number of petals of $P$ is equal to the number of petals of $Q$, as in the above labeling. Two petals $X_i$ and $Y_j$ are said to be *incompatible* if there are vertices $x_i \in C \cap X_i$, and $y_j \in D \cap Y_j$, such that $x_i y_j \notin E$; that is, the vertices of $X_i$ and $Y_j$ cannot be members of the same clique. We also say that two such vertices are incompatible.

We construct a convex geometric graph $H = (B \cup R, F)$ as follows: $B$ and $R$ are sets of $k$ blue (resp. red) points each, placed equidistantly and alternating on some circle $\Omega$ (i.e., $B \cup R$ is the vertex set of a regular $2k$-gon); refer to Fig. 2(middle). The blue points correspond to the petals of $P$, while the red points correspond to the petals of $Q$, and they are labeled in the same way. A blue and a red point are connected by an edge in $F$ if their corresponding petals are incompatible.

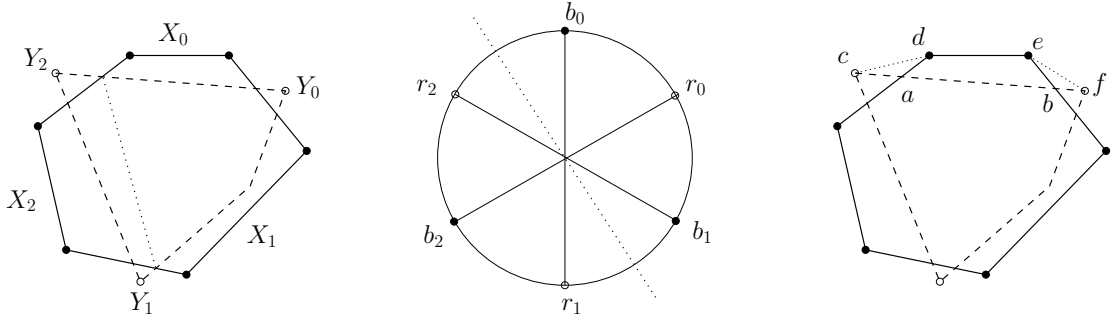

Figure 2: (a) Two overlapping cliques. (b) A hypothetical graph $H$. (c) Illustration to the proof of Lemma 5: $P$ is shown in solid edges, and $Q$ in dashed edges.

**Lemma 5.** *$H$ has no isolated vertices.*

*Proof.* For contradiction, assume that $b_i$ be a blue isolated vertex (the argument is the same for red vertices). The corresponding petal $X_i$ of $P$ is compatible with all petals of $Q$, thus $\tilde{D} = D \cup (X_i \cap S)$ is a clique in $G$. Obviously $\tilde{C} = C \setminus (X_i \cap S)$ is also a clique. Observe that $\tilde{C} \cup \tilde{D} = C \cup D$, that is, $\tilde{C}$ and $\tilde{D}$ cover all the points in $C$ and $D$. It remains to show that $\Psi(\tilde{C}) + \Psi(\tilde{D}) < \Psi(C) + \Psi(D)$, which will contradict the minimality of $\Psi(\mathcal{C})$.

Assume that $a$ and $b$ are the intersection points of $P$ and $Q$ at petal $X_i$, and that $cd$ and $ef$ are the common tangents to $P$ and $Q$ for the pairs of petals $X_i, Y_i$ and $X_i, Y_{i-1}$; see Fig. 2(right). With the notation from the figure, it is enough to check that

$$[|ab|] + [|cd| + |de| + |ef|] < [|ad| + |de| + |eb|] + [|ac| + |ab| + |bf|].$$

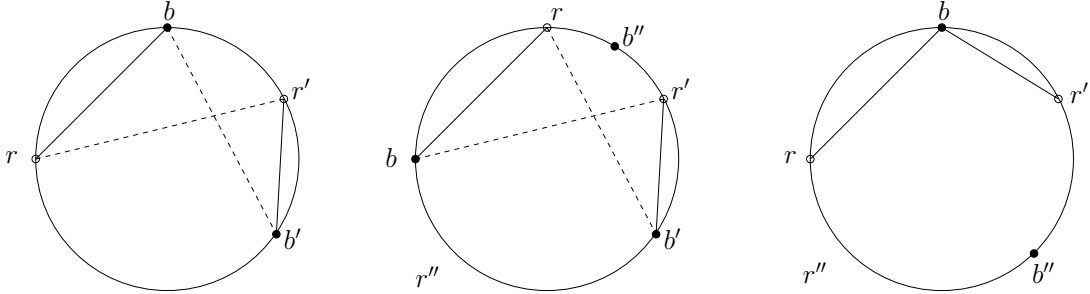

Figure 3: Illustration to the proof of Lemma 6. Blue vertices are drawn as filled circles, red vertices are drawn as empty circles.

This follows immediately from the triangle inequality in $\triangle apq$ and $\triangle bp'q'$, namely $|cd| < |ac| + |ad|$, and $|ef| < |be| + |bf|$. □

Two disjoint edges of $H$ are said to be *anti-parallel* if the 4 endpoints have alternating color around the circle, and *parallel* otherwise.

**Lemma 6.** *$H$ has no pair of disjoint edges.*

*Proof.* We first show that $H$ has no pair of anti-parallel edges; refer to Fig. 3(left). Assume that $br$ and $b'r'$ are two anti-parallel edges. Let $X$ and $X'$ be the petals of $P$ corresponding to $b$ and $b'$, and $Y$ and $Y'$ be the petals of $Q$ corresponding to $r$ and $r'$. Let $x \in X$, $x' \in X'$, $y \in Y$, $y' \in Y'$ be 4 points of $S$. By the triangle inequality, we have

$$2 < |xy| + |x'y'| \leq |xx'| + |yy'| \leq 2,$$

a contradiction.

We now show that $H$ has no pair of parallel edges; refer to Fig. 3(middle). Assume that $br$ and $b'r'$ are two parallel edges. Since the colors of vertices of $H$ are alternating along the circle, there exists a blue vertex, say $b''$, between $r$ and $r'$. By Lemma 5, there exists a red vertex $r''$, so that $b''r''$ is an edge of $H$. It is easy to see that regardless on its position on the circle, either $br$ and $b''r''$ are anti-parallel, or $b'r'$ and $b''r''$ are anti-parallel, contradicting the first part of the proof. □

**Lemma 7.** *Every vertex of $H$ has degree one.*

*Proof.* Assume that $b$ is a blue vertex with (at least) two incident edges $br$ and $br'$; see Fig. 3(right). As in the proof of Lemma 6, there exists a blue vertex, say $b''$, between $r$ and $r'$, and a red vertex $r''$, so that $b''r''$ is an edge of $H$ ($r''$ could be $r$ or $r'$). Regardless on its position on the circle, either $br$ and $b''r''$ are anti-parallel, or $br'$ and $b''r''$ are anti-parallel, contradicting Lemma 6. □

By the three previous lemmas, $H$ consists of a red-blue perfect matching $M$ of $k$ pairwise crossing segments. Moreover, $k$ must be odd, and this matching is unique: $M = \{b_i r_{i+\lfloor k/2 \rfloor} : i = 0, \ldots, k-1\}$, as in Fig. 2(middle). Observe that any halving line of the $2k$ points in $B \cup R$ cuts all edges of $M$ (i.e., all edges of $H$); take for instance the halving line separating $b_0$ from $r_{\lfloor k/2 \rfloor}$, and the corresponding chord $s$ of $I = P \cap Q$ connecting the two intersection points of $P$ and $Q$, as in Fig. 2(left). Let $\tilde{C}$ and $\tilde{D}$ be be the set of points of $C \cup D$ left and respectively right of the cutting line through $s$. Let $\tilde{P} = \text{conv}(\tilde{C})$, and $\tilde{Q} = \text{conv}(\tilde{D})$. Since all edges of $H$ are cut, there is no incompatible pair of vertices in $\tilde{C}$ or in $\tilde{D}$, thus both are cliques. To finish the proof of Theorem 1, it remains to show that $\Psi(\tilde{C}) + \Psi(\tilde{D}) < \Psi(C) + \Psi(D)$, which will contradict the minimality of $\Psi(\mathcal{C})$. We make use of the following simple but useful fact, noted in [5, 22]:

$$\text{per}(P) + \text{per}(Q) = \text{per}(L) + \text{per}(I) = \text{per}(\tilde{P}) + \text{per}(\tilde{Q}) + \text{per}(I) - 2|s|. \tag{9}$$

Since $\text{area}(I) > 0$, we have $\text{per}(I) > 2|s|$, hence

$$\Psi(\tilde{C}) + \Psi(\tilde{D}) = \text{per}(\tilde{P}) + \text{per}(\tilde{Q}) < \text{per}(P) + \text{per}(Q) = \Psi(C) + \Psi(D),$$

as claimed. It follows that $\mathcal{C}$ has no overlapping cliques, and the proof of Theorem 1 is complete.